\documentclass{article}

\usepackage[left=1in, right=1in, top=1in, bottom=1in]{geometry}
\usepackage{graphicx}
\usepackage{color}
\usepackage{soul}
\usepackage[hidelinks]{hyperref}
\usepackage{cleveref}
\usepackage{authblk}
\usepackage{url}

\newif\iflongversion
\longversiontrue

\title{What is Quantum Computer Security?}

\author{Sanjay Deshpande and Jakub Szefer}
\affil{Computer Architecture and Security Lab, Dept. of Electrical and Computer Engineering, Northwestern University, Evanston, IL, USA}

\date{October 2025}

\begin{document}

\maketitle

\iflongversion

\noindent \emph{ This article is a slightly extended version of a one-page SIGDA Newsletter article of the same title that is scheduled to be published in Fall 2025.}

\fi

\iflongversion

\begin{figure}[!h]
    \centering
    \includegraphics[width=1\linewidth]{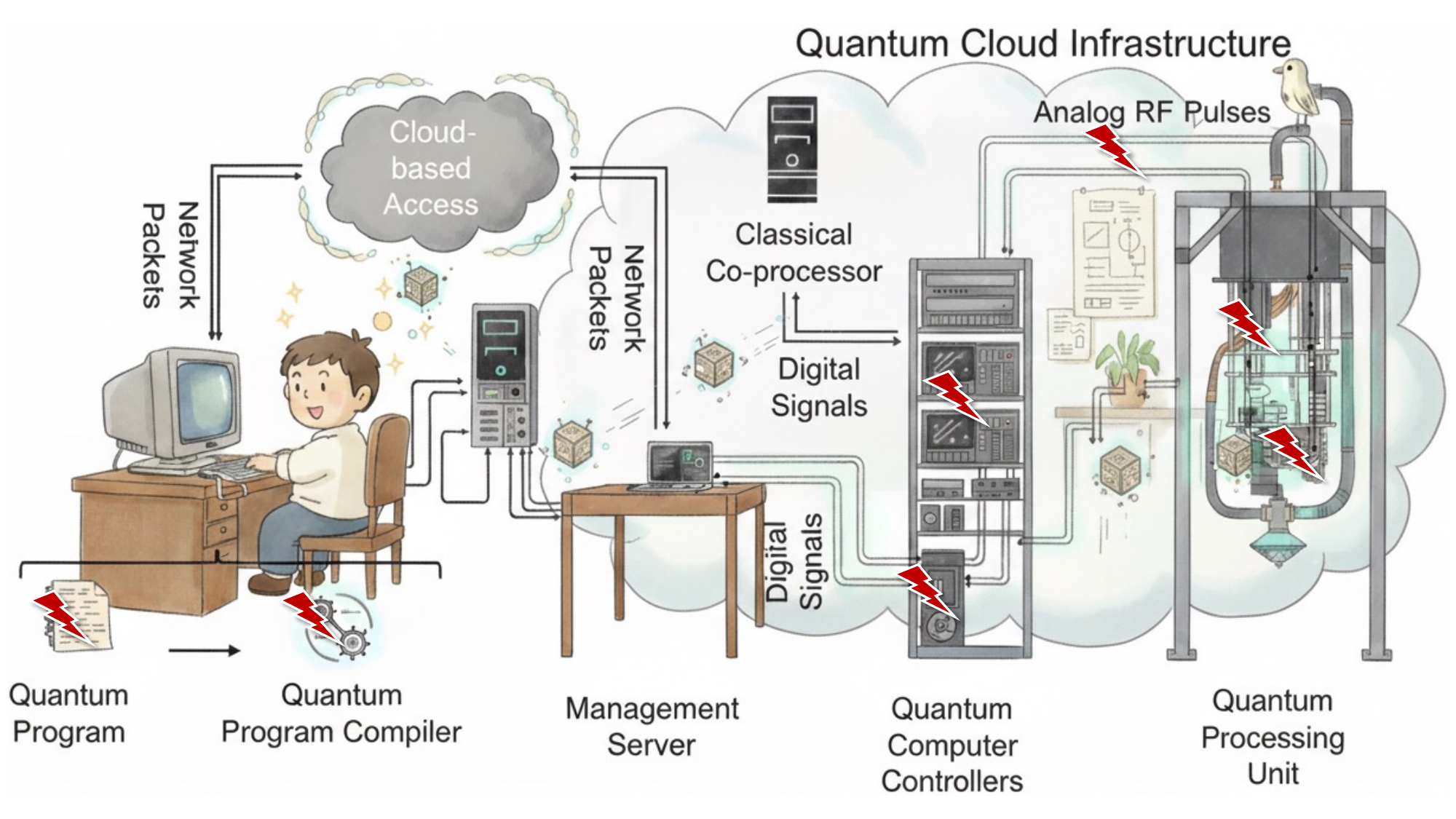}
    \caption{Diagram of a typical cloud-based superconducting quantum computer components and workflow. The attack surface with major potential attack points is superimposed on the diagram.}
    \label{fig:qc_setup}
\end{figure}

\fi

Quantum computing is rapidly emerging as one of the most transformative technologies of our time. With the potential to tackle problems that remain intractable for even the most powerful classical supercomputers, quantum hardware has advanced at an extraordinary pace~\cite{buchs2025rolequantumcomputingadvancing}. Today, major platforms such as {IBM Quantum}, {Amazon Braket}, and {Microsoft Azure} provide cloud-based access to quantum processors, making them more widely available than ever before. While a promising technology, quantum computing is not magically immune to security threats. Much research has been done on post-quantum cryptography, which addresses how to protect classical computers from attackers using quantum computers. This article meanwhile introduces the dual idea of quantum computer security: how to protect quantum computers from security attacks.

\iflongversion

Quantum computers do have some security benefits. One of the security advantages of quantum systems is rooted in the so-called no-cloning theorem, which prohibits the exact copying of unknown quantum states~\cite{Park1970TheCO}. This fundamental property removes an entire class of security attack vectors common in classical computing. For instance, techniques such as buffer overflow exploits, which often rely on the ability to replicate memory contents are not directly applicable in quantum systems. However, while quantum mechanics and no-cloning theorem offers a layer of security, there are other, entirely new sets of vulnerabilities that threaten quantum computers.

Unlike classical processors, quantum computers demand an extensive control infrastructure. To execute quantum operations, specialized classical components, including signal generators, mixers, and FPGAs, must work in concert. A diagram of a typical quantum computer setup is shown in \Cref{fig:qc_setup}. These devices have not been studied comprehensively from a security perspective, yet they represent critical points of potential compromise. A further complication stems from the physical scale of current quantum systems: housed in large server racks, they are inherently more accessible than the microchip-scale architecture of classical CPUs. This accessibility makes them more vulnerable to physical probing and side-channel attacks, particularly targeting the classical control pathways.  

Current systems also lack quantum memory and networking capabilities, meaning all data must be embedded in the quantum program itself. On cloud platforms, submitted jobs therefore include hardcoded constants and computational data. If these programs are intercepted or exposed, attackers could extract sensitive information, raising substantial concerns around data confidentiality. These unique challenges open a fertile landscape for further research. 

\fi

\iflongversion As quantum computing transitions from theoretical promise to practical reality, the security landscape of these systems has emerged as a critical research frontier. \fi Emerging research on security of quantum computers already has uncovered vulnerabilities across the quantum computing stack: from physical hardware to software compilers to algorithmic implementations. While simultaneously innovative defense mechanisms to address these threats have been suggested. The security threats and current state of the defenses is discussed below.

\paragraph{Hardware-Level Vulnerabilities}

At the physical layer, researchers have identified quantum crosstalk as a significant attack vector in multi-tenant cloud environments. Foundational characterization studies~\cite{PhysRevApplied.12.064022} have enabled precise understanding of qubit coupling mechanisms, paving the way for subsequent work~\cite{ash2020analysis, choudhury2025crosstalk, Park1970TheCO} demonstrating how malicious circuits can degrade the fidelity of victim computations sharing the same quantum processor. Beyond crosstalk, side-channel attacks present equally concerning vulnerabilities. Researchers have shown that system behaviors leak sensitive information through multiple channels: timing analysis of reset operations reveals program execution patterns~\cite{mi2022securing}, while power consumption traces enable reverse-engineering of gate-level circuits~\cite{xu2023exploration}. Perhaps most troubling, recent work~\cite{10.1145/3576915.3623104} revealed that standard reset gates fail to properly clear qubit states, allowing information to leak between consecutively executed circuits, a fundamental flaw in quantum system isolation.

\paragraph{Software-Layer Threats}

Moving up the stack, quantum compiler security has emerged as another critical concern. Malicious or compromised compilers pose direct threats to intellectual property, with demonstrated capabilities for circuit theft~\cite{suresh2021short}. More sophisticated attacks exploit the compilation process itself: the QTrojan attack~\cite{Chu:2023lvl} demonstrates how adversaries can stealthily disable data encoding by manipulating hardware configuration files while disguising these modifications as routine pulse calibrations.

\paragraph{Building Defense-in-Depth}

The quantum computer security community is actively developing countermeasures across multiple layers. Hardware-focused defenses include Quantum Trusted Execution Environments~\cite{trochatos2023quantum}, Quantum Physical Unclonable Functions (QPUFs)~\cite{pirnay2022learning,phalak2021quantum}, device fingerprinting~\cite{smith2023fast}, and circuit watermarking~\cite{9424311}. To protect intellectual property in cloud deployments, researchers have proposed circuit obfuscation techniques~\cite{suresh2021short,trochatos2024dynamic} that preserve functionality while obscuring implementation details. Software defenses complement these hardware protections. Quantum antivirus systems~\cite{deshpande2023design} utilize subgraph isomorphism detection to identify malicious circuit patterns that can induce crosstalk. Circuit splitting approaches~\cite{saki2021split} distribute computations across multiple quantum processors to limit information exposure to any single provider.

\paragraph{Taxonomizing the Threat Landscape}

To organize this rapidly evolving field, recent survey work~\cite{chen2024nisq} proposes a systematic taxonomy categorizing quantum security threats into three classes: (1) \textit{Information Leak} attacks that exploit vulnerabilities to extract sensitive data, (2) \textit{Untargeted Attacks} that opportunistically degrade system performance without specific objectives, and (3) \textit{Targeted Attacks} that leverage deep system knowledge to achieve precise malicious goals. Community resources~\cite{qc-hardware-cybersecurity-bibtex} maintain comprehensive bibliographies tracking this literature.

\paragraph{The Path Forward}

As quantum systems grow in capability and accessibility, security cannot be an afterthought. The research community must continue developing robust defenses that evolve alongside quantum technology itself. Only through sustained attention to security, spanning hardware design, software toolchains, and algorithmic implementations, can we ensure that quantum computing delivers not just computational power, but trustworthy, resilient systems worthy of handling user's most sensitive computations.

\small

\bibliographystyle{acm}
\bibliography{refs}

\end{document}